\documentclass[prapplied,aps,superscriptaddress,showpacs,reprint]{revtex4-2}

\usepackage{graphicx}
\usepackage{dcolumn}
\usepackage{bm}
\usepackage{SIunits}
\usepackage{color}
\usepackage{braket}
\usepackage[dvipsnames]{xcolor}
\usepackage[colorlinks=true, citecolor={blue!80!black}, urlcolor={blue!50!black}, linkcolor = {blue!80!black}]{hyperref}
\usepackage{appendix}
\usepackage[symbol]{footmisc}

\usepackage{booktabs}

\usepackage{soul} 
\begin{document}
	
	
	\title{Gate-Tunable Transmon Using Selective-Area-Grown Superconductor-Semiconductor Hybrid Structures on Silicon}
	
	\author{Albert Hertel}
	\altaffiliation{Current address: Institute for Semiconductor Nanoelectronics, Peter Gr\"unberg Institute 9, Forschungszentrum J\"ulich \& J\"ulich-Aachen Research Alliance (JARA), Forschungszentrum J\"ulich and RWTH Aachen University, Germany}
	\thanks{These two authors contributed equally.}

	\author{Michaela Eichinger}
	\thanks{These two authors contributed equally.}
	\affiliation{Center for Quantum Devices, Niels Bohr Institute, University of Copenhagen, 2100 Copenhagen, Denmark}
	
	\author{Laurits O. Andersen}
	\author{David M. T. van Zanten}
	\author{Sangeeth Kallatt}
	\affiliation{Center for Quantum Devices, Niels Bohr Institute, University of Copenhagen, 2100 Copenhagen, Denmark}

	\author{Pasquale Scarlino}
	\affiliation{Microsoft Quantum Lab--Copenhagen, 2100 Copenhagen, Denmark}

	\author{Anders Kringh\o j}
	\author{Jos\'e M. Chavez-Garcia}
	\affiliation{Center for Quantum Devices, Niels Bohr Institute, University of Copenhagen, 2100 Copenhagen, Denmark}
	
	\author{Geoffrey C. Gardner}
	\author{Sergei Gronin}
	\affiliation{Birck Nanotechnology Center, Purdue University, West Lafayette, Indiana 47907 USA}
	\affiliation{Microsoft Quantum Lab--West Lafayette, West Lafayette, Indiana 47907 USA}
	
	\author{Michael J. Manfra}
	\affiliation{Birck Nanotechnology Center, Purdue University, West Lafayette, Indiana 47907 USA}
	\affiliation{Microsoft Quantum Lab--West Lafayette, West Lafayette, Indiana 47907 USA}
	\affiliation{Department of Physics and Astronomy, Purdue University, West Lafayette, Indiana 47907 USA}
	\affiliation{School of Materials Engineering, and School of Electrical and Computer Engineering, Purdue University, West Lafayette, Indiana 47907 USA}

	\author{Andr\'as Gyenis}
	\altaffiliation{Current address: Department of Electrical, Computer \& Energy Engineering, University of Colorado Boulder, CO 80309, USA.}
	\affiliation{Center for Quantum Devices, Niels Bohr Institute, University of Copenhagen, 2100 Copenhagen, Denmark}

	\author{Morten Kjaergaard}	
	\author{Charles M. Marcus}
	\affiliation{Center for Quantum Devices, Niels Bohr Institute, University of Copenhagen, 2100 Copenhagen, Denmark}
	
	\author{Karl D. Petersson}
	\affiliation{Center for Quantum Devices, Niels Bohr Institute, University of Copenhagen, 2100 Copenhagen, Denmark}
	\affiliation{Microsoft Quantum Lab--Copenhagen, 2100 Copenhagen, Denmark}
	
	\date{\today}

	\begin{abstract}
		We present a gate-voltage tunable transmon qubit (gatemon) based on planar InAs nanowires that are selectively grown on a high resistivity silicon substrate using\linebreak III-V buffer layers. We show that low loss superconducting resonators with an internal quality of $2\times 10^5$ can readily be realized using these substrates after the removal of buffer layers. We demonstrate coherent control and readout of a gatemon device with a relaxation time, $T_{1}\approx 700\,\mathrm{ns}$, and dephasing times, $T_2^{\ast}\approx 20\,\mathrm{ns}$ and $T_{\mathrm{2,echo}} \approx 1.3\,\mathrm{\mu s}$. Further, we infer a high junction transparency of $0.4 - 0.9$ from an analysis of the qubit anharmonicity. 
		
	\end{abstract}
	
	\maketitle

	\section{\label{sec:level1}Introduction}
	
	Superconducting qubit systems are one of the most promising approaches towards realizing a large scale fault tolerant quantum processor \cite{Arute2019}.~In recent years, a transmon variant that uses voltage-tunable superconductor-semiconductor hybrid Josephson junctions (JJs), the gatemon, has been developed \cite{Larsen2015, deLange2015} and realized in several material systems such as vapor-liquid-solid (VLS) nanowires \cite{Larsen2015}, two-dimensional electron gases (2DEG) \cite{Casparis2018} and graphene \cite{Wang2019}. These junctions have an inherent non-sinusoidal current-phase relationship and could be used to create new building blocks for quantum computing or as a tool for condensed matter experiments based on cQED techniques \cite{Kringhoej2018, Wang2019, Kringhoej2020, Larsen2020}. Further, these qubits could be interfaced with low dissipation cryogenic CMOS control systems \cite{Pauka2021}. 
	
	In this work, we demonstrate coherent operations of a gatemon qubit fabricated using a material system with selective-area-grown Al-InAs hybrid structures \cite{Krizek2018, Hertel2021} on a Si substrate that has planar III-V buffer layers (Si SAG). This material system has the potential to combine the individual advantages of previously demonstrated material systems \cite{Larsen2015, Casparis2016, Casparis2018, Wang2019} for future gatemon devices. First, similar to 2DEG-based gatemons \cite{Casparis2018}, the number of JJs are relatively easy to scale as the Al-InAs nanowires are monolithically integrated into the substrate using deterministic lithographic patterning techniques. Second, the qubit capacitor, readout and control components can be fabricated directly on the high resistivity silicon substrate with low dielectric loss \cite{Wang2015}, similar to VLS-nanowire gatemons \cite{Larsen2015}. Similar experiments implemented on III-V substrates report a resonator quality factor of $6 \times 10^4$ \cite{Casparis2018} or lower \cite{Toida2013, Scigliuzzo2020, Yuan2021, Casparis2018} and qubit relaxation times up to a few microseconds \cite{Casparis2018}. The main source of losses on III-V substrates can be explained by the piezoelectric photon-phonon coupling in these materials \cite{Scigliuzzo2020, Toida2013}.
	
	\begin{figure}[h!]
		\includegraphics[width=1\columnwidth]{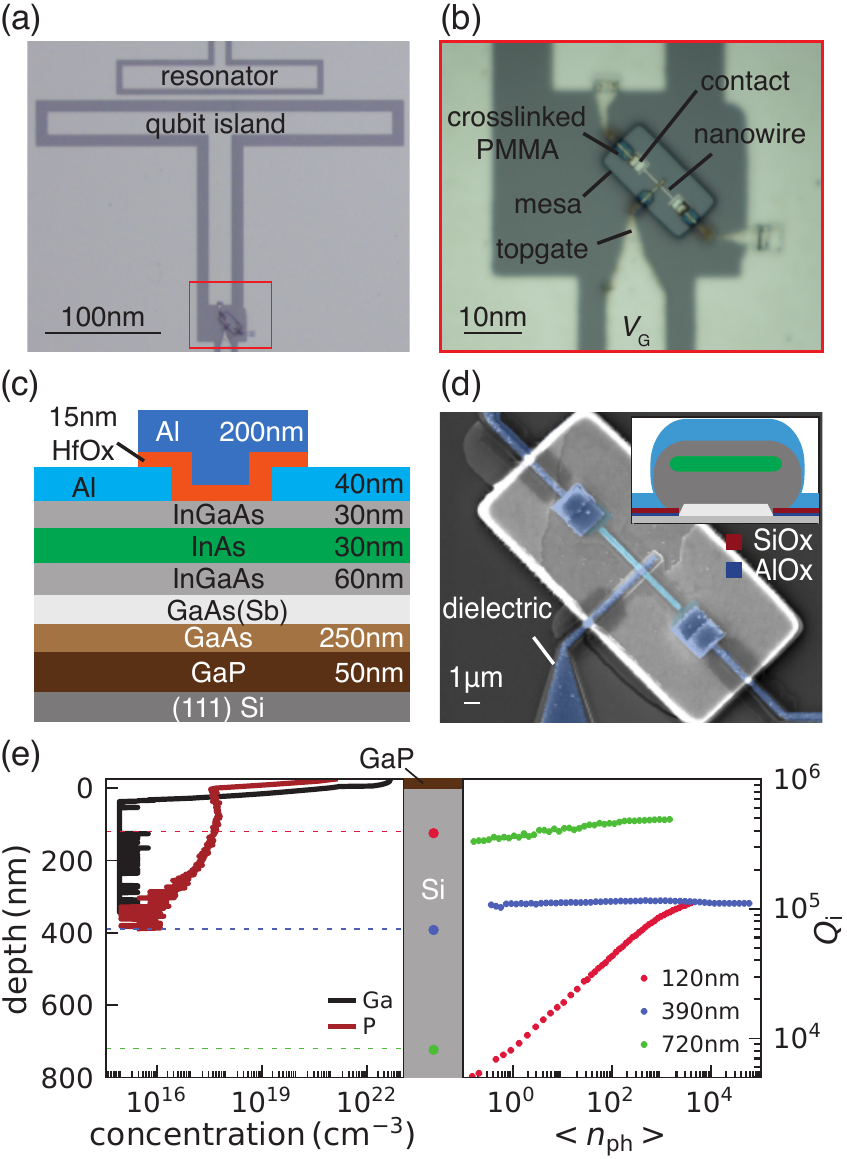}
		\setlength\belowcaptionskip{-12pt}
		\caption{\label{fig:Fig1} \textbf{Qubit devices and microwave properties}. \textbf{(a)} Optical micrograph of the T-shaped qubit island that is capacitively coupled to a readout cavity. (\textbf{b)} Zoomed in optical micrograph of the junction region. The junction is positioned on a~$\sim1\,\mathrm{\micro m}$ thick mesa. \textbf{(c)} Schematic of material stack around the junction. The InAs region (green) is proximitized by the epitaxial Al (light blue) and the junction is formed by selectively removing a short segment of the Al. \textbf{(d)} Scanning electron micrograph of the junction area. The mesa consists of GaAs, GaP and $400\,\mathrm{nm}$ of Si. The top gate metal climbs the mesa with the help of crosslinked PMMA. (inset) Cross section of the nanowire with the dielectric growth mask consisting of SiOx and AlOx. \textbf{(e)} (Left) Ga and P concentration of the material stack as a function of depth measured with secondary ion mass spectrometry. (Center) Schematic of the corresponding material stack. (Right). Internal quality factor of test resonators as a function of photon number for an etch depth into the Si of $120\,\mathrm{nm}$ (red), $390\,\mathrm{nm}$ (blue) and $720\,\mathrm{nm}$ (green).  }
	\end{figure}
	
	In this study, we demonstrate that the material system can be used to make gatemons. We introduce the device design and show that high quality resonators can be fabricated with the material system after the removal of the buffer layers. Next we show gate tunable qubits and study the qubit anharmonicity, concluding a high junction transparency. Finally, we demonstrate coherent oscillations and extract qubit relaxation and dephasing times of a qubit device.
	
	\section{\label{sec:Device}Devices}
	
	Figure~\ref{fig:Fig1}a shows an optical micrograph of the gatemon device. The T-shaped qubit island is capacitively coupled to a quarter-wave coplanar waveguide cavity with resonance frequency $f_{\mathrm{r}}\approx6.6\,\mathrm{GHz}$ and has a total charging energy $E_{\mathrm{C}}/h \approx 225\,\mathrm{MHz}$ extracted from finite-element simulations. All parts of the readout circuit such as the cavity, transmission line and the qubit island are fabricated on the high resistivity (111) Si substrate with a $4\degree$ miscut after etching the III-V layers. The qubit island is connected to the ground plane through a JJ that is located on a mesa with dimensions $\mathrm{10\,\micro m\,x\,20\,\micro m}$ (Fig.~\ref{fig:Fig1}b). The JJ is formed by selectively wet etching a $\sim120\,\mathrm{nm}$ long segment of the Al on the $\sim300\,\mathrm{nm}$ wide InAs SAG nanowire. The Josephson energy $E_{\mathrm{J}}(V_{\mathrm{G}})$ of the junction is controlled by applying a gate voltage $V_{\mathrm{G}}$ via the Al top gate which is separated from the junction by $15\,\mathrm{nm}$ thick gate dielectric ($\mathrm{HfO_2}$). The qubit frequency is given by $f_{\mathrm{q}}\approx\sqrt{8E_{\mathrm{J}}(V_{\mathrm{G}})E_{\mathrm{C}}}/h$ in the transmon limit ($E_{\mathrm{J}}\gg E_{\mathrm{C}}$) \cite{Koch2007}. The top gate and Al wires that connect the nanowire with the ground plane and qubit island climb the mesa aided by layers of crosslinked polymethyl methacrylate resist (PMMA). The primary purpose of the PMMA layers is to decouple the qubit from the lossy III-V material of the mesa. Figure~\ref{fig:Fig1}c shows a schematic of the junction region of the heterostructure along the nanowire. A schematic cross section of the nanowire is shown in the inset of Fig.~\ref{fig:Fig1}d. Further details of the material stack, including electrical transport characterization, are presented in Ref.~\cite{Hertel2021} and briefly summarized here. The  GaP/GaAs buffer is grown using metallic organic chemical vapor epitaxy and used to bridge the lattice mismatch between Si and InAs. Selective area growth \cite{Krizek2018} was used to grow planar InAs nanowires on the GaAs surface by molecular beam epitaxy \cite{Gardner2016}. To define the nanowire regions, thin films of AlOx and SiOx were deposited globally on the wafer and patterned using electron-beam lithography and a combination of reactive-ion etching (RIE) and wet etching techniques. Inside the mask opening the InAs nanowires were grown, where the Sb-dilute GaAs buffer and InGaAs layers are used to improve the InAs quality by promoting strain relaxation \cite{Krizek2018}. The top InGaAs layer is used to prevent surface damage of the InAs layer due to device processing. A blanket Al layer was deposited in-situ to ensure a high quality interface between the Al and semiconductor heterostructure \cite{Krogstrup2015, Chang2015}. 
	
	To fabricate qubit devices most of the epitaxial Al and dielectric growth mask were removed in a first step, followed by etching the GaAs/GaP buffer and $\sim 400\,\mathrm{nm}$ of Si to define mesa structures. The removal of the top layers of Si are necessary for the fabrication of high quality resonators as will be discussed in Section \ref{sec:Microwave_loss}. A detailed summary of the device fabrication can be found in Appendix \ref{app:Fab_qubit}. 
	
	\section{\label{sec:RF-properties}Microwave loss}\label{sec:Microwave_loss}
	
	Figure~\ref{fig:Fig1}e (left panel) shows the doping profile of a Si substrate with GaP and GaAs grown on top of it measured by secondary ion mass spectrometry. The phosphorus atoms diffuse into the Si substrate during the growth process. A P concentration of $10^{16}\,-\,4\times10^{17}\,\mathrm{atoms/cm^{-3}}$ is measured in the top $300\,\mathrm{nm}$ of the Si. The gallium atoms do not diffuse far into the Si and the concentration drops below the detection limit ($\sim 10^{15}\mathrm{cm^{-3}}$) within the first $20\,\mathrm{nm}$. P atoms inside the Si crystal can be considered two-level systems (TLS) \cite{Anderson1972, Phillips1987} which are one of the main limitations for the quality factor of resonators \cite{Gao2008} and the coherence times of superconducting qubits \cite{Lisenfeld2019}.
	
	To investigate the loss due to the P doping, we fabricated test resonators on the Si substrate after removing the III-V layers and further etching into the Si substrate  (see Appendix \ref{app:Fab_res} for further details). The extracted internal quality factor $Q_{\mathrm{i}}$ as a function of the mean photon number $\braket{n_{\mathrm{ph}}}$ for three different etch depths into the Si substrate is shown in the right panel of Fig.~\ref{fig:Fig1}e. For an etch depth of $120\,\mathrm{nm}$ we find $Q_{\mathrm{i}}<10^4$ in the one photon regime. In contrast, we find $Q_{\mathrm{i}} \approx 1\times10^5$ and $Q_{\mathrm{i}} \approx 3\times10^5$ for etch depths $390\,\mathrm{nm}$ and $720\,\mathrm{nm}$, respectively. The dielectric loss tangent associated with these quality factors is low enough to enable transmon devices with lifetimes of several microseconds where the exact limit depends on the qubit geometry due to the participation ratio of the electrical field at the surface and interfaces \cite{Lisenfeld2019, Wang2015, Sandberg2021}. It should be noted that we measured $Q_{\mathrm{i}} \approx 2\times10^5$ for a designated test resonator on the qubit device chip (Figs.~\ref{fig:Fig1}a-b,d) after the entire fabrication.
	
	\section{\label{sec:spectroscopy} Qubit spectroscopy}
	
	\begin{figure}[t]
		\includegraphics[width=1\columnwidth]{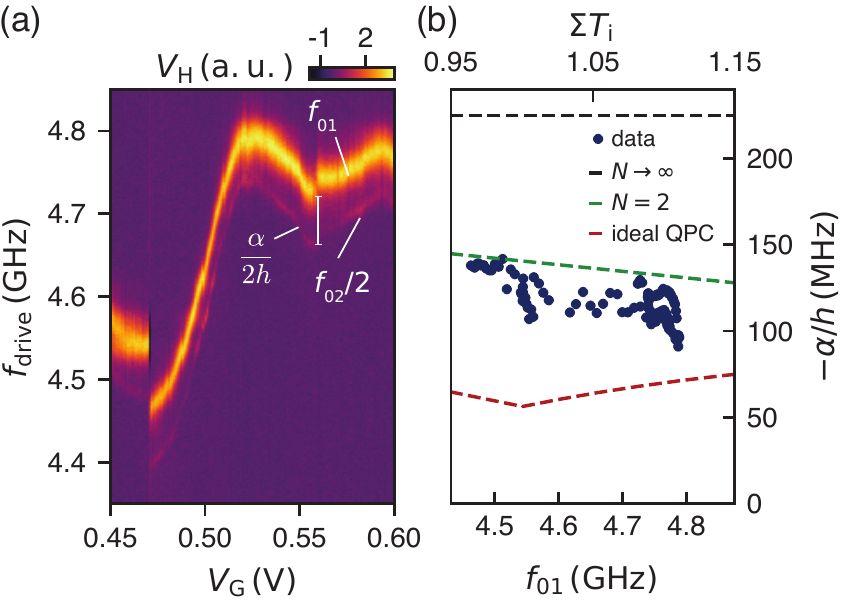}
		\setlength\belowcaptionskip{-8pt}
		\caption{\label{fig:Fig2} \textbf{Qubit anharmonicity}.  \textbf{(a)} Resonator response $V_{\mathrm{H}}$ as a function of gate top gate voltage $V_{\mathrm{G}}$ and drive frequency $f_{\mathrm{drive}}$. The qubit is driven through the gate line with microwave pulses with a relatively high power $P_{\mathrm{rf}}$ on-chip such that the $|0\rangle\rightarrow|1\rangle$ with frequency $f_{01}$ and the $|0\rangle\rightarrow|2\rangle$ with frequency $f_{02}/2$ are measured. \textbf{(b)}. Extracted anharmonicity $-\alpha/h$ as  a function of qubit frequency $f_{\mathrm{01}}$ and $\Sigma T_{\mathrm{i}} = (hf_{01})^2/(2\Delta E_{\mathrm{C}})$.  Dashed lines indicate model prediction for an ideal QPC (red), two channels with equal transmission (green) and the tunneling limit (black).}
	\end{figure}
	
	We next demonstrate gate control of the qubit device (Figs.~\ref{fig:Fig1}a,b,d) and study the transport across the JJ using the qubit anharmonicity, for measurement details see Appendix \ref{app:Measurement_qubit}. Figure~\ref{fig:Fig2}a shows a two-tone spectroscopy measurement of the qubit, where we measured the resonator response $V_{\mathrm{H}}$ as a function of gate voltage $V_{\mathrm{G}}$ and drive frequency $f_{\mathrm{drive}}$. Due to the relatively high on-chip drive power $P_{\mathrm{rf}}\approx-95\,\mathrm{dBm}$ both the $\ket{0}\rightarrow \ket{1}$ transition at frequency $f_{01}$ and two-photon $\ket{0}\rightarrow\ket{2}$ transition at frequency $f_{02}/2$ are resolved. With decreasing $V_{\mathrm{G}}$ the overall qubit frequency decreases and exhibits a non-monotonic gate response that is typically observed for gatemon qubits \cite{Larsen2015,Kringhoej2018, Casparis2018, Wang2019}. We attribute the jumps in $f_{01}$ and $f_{02}/2$ to switches of the channel configuration in the nanowire.
	
	The current-phase relation of a semiconductor-based JJ is non-sinusoidal and can be described with a set of channel transmission coefficients $\{T_{\mathrm{i}}\}$ consisting of a few channels with high transmission eigenvalues \cite{Spanton2017, Goffman2017, Nichele2021}. As a consequence, the gatemon anharmonicity, $\alpha/h =2 (f_{02}/2-f_{01})$, in a gatemon is typically lower than for a metallic transmon \cite{Kringhoej2018}, where $\alpha \approx-E_{\mathrm{C}}$ \cite{Koch2007, Krantz2019}. Here, we analyze $\alpha$ following Ref.~\cite{Kringhoej2018} to study the JJ transmission. We calculate $\{T_{\mathrm{i}}\}$ for each value of gate voltage using $\Sigma T_{\mathrm{i}} = (hf_{01})^2/(2\Delta E_{\mathrm{C}})$, where we use the measured $\Delta/e = 190\,\mathrm{\micro eV}$ (Ref. \cite{Hertel2021}) and $E_{\mathrm{C}}/h = 225\,\mathrm{MHz}$ from finite-element simulations on an intrinsic Si substrate. Assuming $N$ transmitting channels with equal transmission probability, $T$, the model gives $\alpha = -E_{\mathrm{C}}(1-3E_{\mathrm{J}}/(\Delta N))$. Figure~\ref{fig:Fig2}b shows the anharmonicity as a function of $f_{01}$ and $\Sigma T_{\mathrm{i}}$. The data lie between the predicted value for $N=2$ channels and the ideal quantum point contact (QPC) model in which channels are filled in a step-like manner with at most one partially transmitting channel \cite{Beenaker1992}. The data suggest that the transport in the observed gate range is dominated by 2 channels. Given equal transmission probabilities \cite{Kringhoej2018} this sets a lower bound $T_{\mathrm{min}}>\Sigma T_{\mathrm{i}}/2 \sim 0.48$ on the junction transparency. Similar studies on gatemons based on VLS InAs/Al nanowires with in-situ grown epitaxial Al (Ref. \cite{Kringhoej2018}), where $\alpha \approx 100-150\,\mathrm{MHz}$, report two or three channels with $T_{\mathrm{min}} = 0.4 - 0.9$. This result is in good agreement with DC transport data obtained for a nominally identical material stack in Ref. \cite{Hertel2021}, where we conclude a high junction transparency from measurements of the excess current and observed signatures of multiple Andreev reflections.
	
	\section{\label{sec:Coherent_oscillations} Coherent oscillations and Coherence Times}
	
	\begin{figure}[t]
		\includegraphics[width=1\columnwidth]{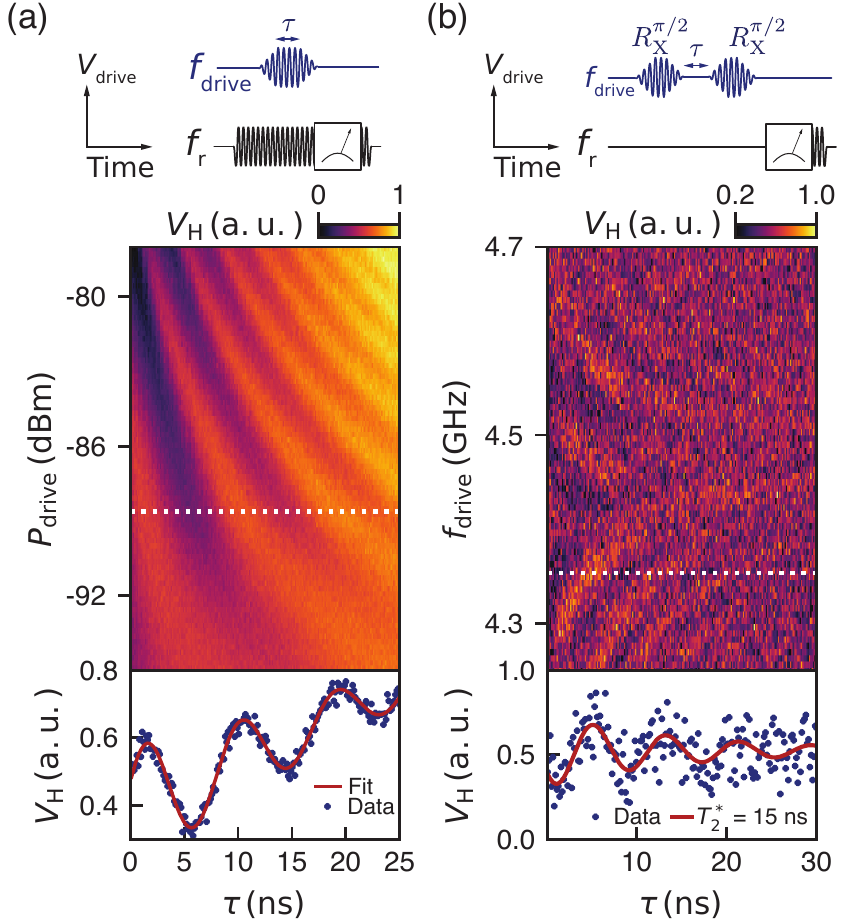}
		\setlength\belowcaptionskip{-16pt}
		\caption{\label{fig:Fig3} \textbf{Coherent oscillations}. \textbf{(a)} The main panel shows Rabi oscillations as a function of drive power on-chip $P_{\mathrm{drive}}$ and pulse length $\tau$ at fixed drive frequency $f_{\mathrm{drive}}=4.47\,\mathrm{GHz}$. The pulse sequence used is shown above the plot. The lower panel shows a line cut at $P_{\mathrm{drive}}=-89\,\mathrm{dBm}$ (white dotted line) from the main panel. The solid line is a fit to the data using a damped sinusoid with a linear contribution. \textbf{(b)} The main panel shows Ramsey fringes as function of delay $\tau$ between $R^{\pi/2}_{\mathrm{X}}$ pulses and $f_{\mathrm{drive}}$. The pulse sequence is shown above. The lower panel shows a line cut at $f_{\mathrm{drive}} = 4.205\,\mathrm{GHz}$ (white dotted line) with a fit to a damped oscillation yielding $T_2^{\ast} \approx 15\,\mathrm{ns}$ }
	\end{figure}
	
	Next, we demonstrate basic qubit control using time-domain manipulation. Figure~\ref{fig:Fig3}a shows coherent Rabi oscillations with a fixed drive frequency $f_{\mathrm{drive}} = f_{\mathrm{q}} = 4.47\,\mathrm{GHz}$ where the on-chip drive power $P_{\mathrm{rf}}$ and the length of the qubit drive pulse $\tau$ were varied. The lower panel of Fig.~\ref{fig:Fig3}a shows the fit to the data at $P_{\mathrm{drive}}=-89\,\mathrm{dBm}$, where the data are described by a damped sinusoid with a linear increase. This additional linear increase in $V_{\mathrm{H}}$ (Fig.~\ref{fig:Fig3}b) could be the result of leakage to higher order states \cite{Peterer2015}. This leakage is expected as the Rabi frequency is similar to the anharmonicity $|\alpha|/h$. See Appendix \ref{app:Measurement_qubit} for further details. 
	
\begin{figure}[t]
		\includegraphics[width=1\columnwidth]{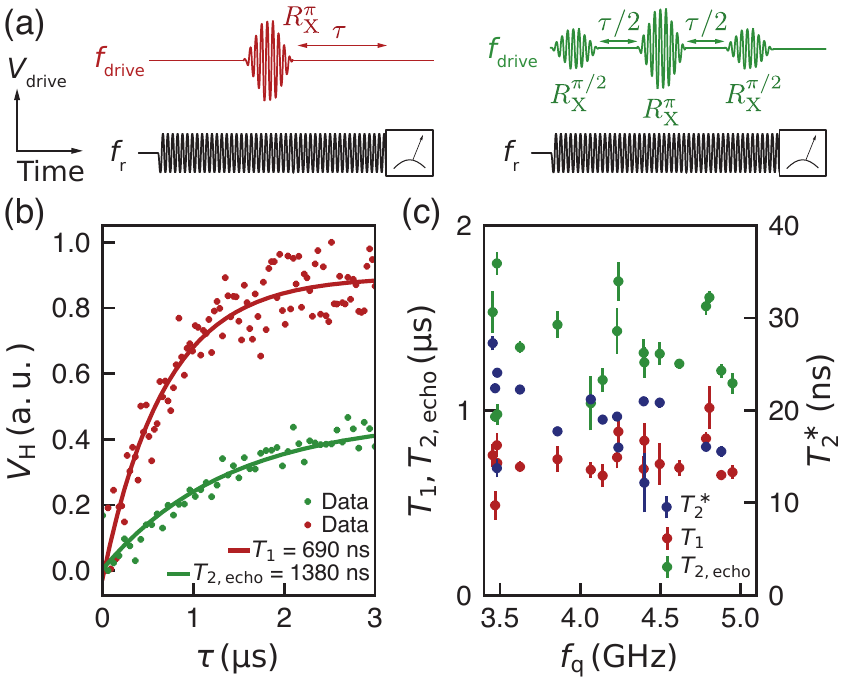}
		\setlength\belowcaptionskip{-8pt}
		\caption{\label{fig:Fig4} \textbf{Qubit lifetimes}. \textbf{(a)} Pulse sequence for lifetime measurements. \textbf{(b)} $T_1$ and $T_{\mathrm{2,echo}}$ measurement of the qubit at $f_{\mathrm{q}}\approx4.48\,\mathrm{GHz}$. \textbf{(c)} Lifetimes as a function of $f_{\mathrm{q}}$. $T_1$ and $T_{\mathrm{2,echo}}$ were measured with the pulse sequence shown in (a). $T_{\mathrm{2}}^{\ast}$ was extracted from the power dependence of the linewidth of the qubit in spectroscopy (Appendix \ref{app:T2star}).}
	\end{figure}

	Figure \ref{fig:Fig4}b shows Ramsey fringes measured by using two $R^{\pi/2}_{\mathrm{X}}$ pulses separated by a delay $\tau$. The lower panel of the figure shows a fit of a damped sinusoid to a linecut of the data. From the fit we extract a dephasing time $T_2^{\ast} \approx 15\,\mathrm{ns}$. The extracted value is in good agreement with estimates of $T_2^{\ast}$ from a fit to the power dependence of the qubit linewidth in spectroscopy as described in Ref. \cite{Schuster2005} (see Appendix \ref{app:T2star}). 

	Figure~\ref{fig:Fig4}a shows the pulse sequences that we use to measure the qubit lifetimes. We use overlapping drive and readout pulses to increase the signal-to-noise-ratio as the $T_2^{\ast}$ time is shorter than the ring-up time of the readout resonator $\tau_{\mathrm{rise}} \approx 480\,\mathrm{ns}$ (see Appendix~\ref{app:Measurement_qubit}). Figure~\ref{fig:Fig4}b shows representative measurements of $T_1$ (red) and $T_{\mathrm{2,echo}}$ (green) using a Hahn echo sequence with fits to an exponential decay. 
	
	Figure~\ref{fig:Fig4}c shows the measured lifetimes for qubit frequencies between $3.5$ and $5.0\,\mathrm{GHz}$. $T_1$ and $T_{\mathrm{2,echo}}$ are obtained from calibrated pulses and $T_2^{\ast}$ is estimated from the qubit linewidth (Appendix \ref{app:T2star}). The extracted mean values are $\overline{T}_1 = (740\pm110)\,\mathrm{ns}$, $\overline{T}_2^{\ast} = (21\pm7)\,\mathrm{ns}$, and $\overline{T}_{\mathrm{2,echo}} = (1340\pm230)\,\mathrm{ns}$. The $\overline{T}_1$ and $\overline{T}_{\mathrm{2,echo}}$ times are comparable with the lifetimes extracted for the first gatemons fabricated with VLS nanowires on a Si substrate \cite{Larsen2015}. However, an order of magnitude longer $T_2^{\ast} \approx 900\,\mathrm{ns}$ were reported for those devices \cite{Larsen2015, Casparis2016}.

	\section{\label{sec:conclusion}Discussion and Conclusions}
	
	The measured dephasing times indicate that dephasing is limited by low-frequency noise since $T_{\mathrm{2,echo}} \gg T_{\mathrm{2}}^{\ast}$. We suspect that both the gate dielectric and dielectric close to the junction used for the selective area growth \cite{Hertel2021} cause the low decoherence times. Comparing previous Al-InAs based gatemon architectures, VLS-nanowire gatemons \cite{Larsen2015, Casparis2016} reported longer dephasing times ($T_{\mathrm{2,VLS}}^{\ast} \approx 0.9 - 3.7\,\mathrm{\micro s}$) than gatemons based on 2DEGs \cite{Casparis2018} $(T_{\mathrm{2,2DEG}}^{\ast} \approx 400\,\mathrm{ns})$ which are longer than the dephasing times in this work ($T_{\mathrm{2,SAG}}^{\ast}\approx 20\,\mathrm{ns}$). For the VLS-nanowire gatemon the nanowire was placed on the device chip without additional dielectrics near the active junction region. In case of the 2DEG-gatemon, mesa material below the proximitized InAs channel and gate dielectric were still present near the junction on the qubit chip after device fabrication. Further, the mesa with a width of $\sim 1\,\mathrm{\micro m}$ was smaller then the mesa in this work ($\sim10\,\mathrm{\micro m}$). The main difference in this work is the presence of residual growth dielectric with a low quality near the junction. The $\mathrm{SiO_x}$ layer in particular was optimized to have a low crystal quality to increase the selectivity of the InAs growth \cite{Krizek2018}. We expect this layer to have an increased TLS density compared to the mesa material and thus limit $T_{\mathrm{2}}^{\ast}$. 
	
	The relaxation times of the qubits are lower than the relaxation times of the test resonator, indicating that $T_1$ is not limited by the dielectric loss of the substrate. Based on the internal quality factor measured for readout resonators $Q_{\mathrm{i}} = 2\times10^5$, we would expect a relaxation time for a qubit with frequency $f_{\mathrm{q}}=5\,\mathrm{GHz}$ of $T_1 \approx Q_{\mathrm{i}}/(2\pi f_{\mathrm{q}})\approx 6\,\mathrm{\micro s}$. Instead the qubit could be limited by dielectric loss due to the III-V material or doped Si in the mesa region. Although we do not expect that the doping level is high enough for the doped Si to be conducting, we cannot exclude the possibility that normal conducting channels exist near the GaP/Si interface.
	
	To improve $T_{\mathrm{2}}^{\ast}$, the material quality near the junction needs to be improved. Further, the qubit can be better decoupled from the mesa by means of device design which leads to a reduced participation ratio \cite{Wang2015, Bilmes2020, Sandberg2021} and, therefore, increased relaxation times. Potential changes in device fabrication and design are summarized in Appendix \ref{app:roadmap}. Other potential sources of decoherence such as gate noise and readout are discussed in Appendix \ref{app:decoherence}. 
	
	In summary, we have demonstrated that selective-area-grown Al-InAs nanowires on a Si substrate are a potential platform for voltage controlled transmon qubits. We implemented the SAG based gatemon on a low-loss substrate with relaxation times of roughly $700\,\mathrm{ns}$ and dephasing times of $20\,\mathrm{ns}$. With further improvement of coherence this material platform opens new possibilities for scalable highly integrated quantum circuits such as gatemons \cite{Larsen2015}, protected qubits \cite{Larsen2020}, voltage-tunable quantum buses \cite{Casparis2019} and voltage-tunable quantum memories \cite{Sardashti2020}.

	\begin{acknowledgments}
		We thank Anna Mukhortova for help with the process development. This work was supported by the Microsoft Corporation, the Danish National Research Foundation and VILLUM FONDEN. 
	\end{acknowledgments}
	
	\begin{appendix}
		\section{Qubit device fabrication}\label{app:Fab_qubit}
		The qubit devices were fabricated using a combination of standard electron-beam lithography and UV lithography. At the beginning of the qubit fabrication, the entire chip was covered with a $40\,\mathrm{nm}$ thick Al film on top of the dielectric layers consisting of $10\,\mathrm{nm}$ thick $\mathrm{SiO_x}$ and $5\,\mathrm{nm}$ thick $\mathrm{AlO_x}$. In a first step, Al was etched selectively on the device chips using Al Etchant Transene D, only leaving Al in dumbbell-shaped areas around nanowires (Fig.~\ref{fig:Fig1}b). Using the same resist stack, both $\mathrm{SiO_x}$ and $\mathrm{AlO_x}$ were etched using buffered HF solution. The dumbbell shape offers a compromise between little dielectric and Al being left around the nanowire [see Fig.~\ref{fig:Fig1}b] and sufficiently large MBE Al patches with an area of $\sim 1\,\mathrm{\micro m}^2$ that are used to connect the Al-InAs nanowires to the rest of the circuit. Next, the mesa was defined by protecting the nanowire and a small surrounding area with photoresist (AZ5214E) while removing GaAs, GaP and $\sim 400\,\mathrm{nm}$ of Si using two reactive ion etching steps. First, an etch with process gases $\mathrm{Cl_2}$ and $\mathrm{Ar}$ was used that created an almost vertical mesa profile. The second etch, which used process gases $\mathrm{Cl_2}$ and $\mathrm{N_2}$, created a trapezoidal mesa profile (see Fig.~\ref{fig:Fig1}d). The first step was optimized to etch the material stack fast, reducing heat load of the resist during the process, and creating steep side walls. The second step creates a trapezoidal shape, which is beneficial for the crawl up of top gate metal. Then, the JJ was defined by selectively removing a $\sim 150\,\mathrm{nm}$ long Al segment on the nanowire using Transene Al Etchant Type D at $50\,\mathrm{^{\circ}C}$. Next, Al for the readout circuit and qubit island definition was evaporated with a lift-off process, where the mesas were protected by resist. The evaporation was preceded by a $10\,\mathrm{s}$ long dip in buffered HF solution to remove surface oxides from the chip. The transmission line, readout resonators, qubit islands, a test resonator and gate lines were defined by selectively removing Al with a wet-etch solution (Transene Al Etchant Type D at $50\,\mathrm{^{\circ}C}$). Next, $15\,\mathrm{nm}$ $\mathrm{HfO}_2$ was grown by atomic layer deposition in lithographically pre-defined regions on top of the JJs. Additionally $\mathrm{HfO}_2$ was deposited in areas where Al would climb mesas to ensure they were isolated. In these climbing areas PMMA bridges were defined by crosslinking PMMA through the exposure with 30 times the area dose that is typically used to pattern the resist (see Fig.~\ref{fig:Fig1}d). Next, gates ($200\,\mathrm{nm}$ thick Al) were evaporated. These were later used to tune the critical currents of the single JJs and thereby the qubit frequencies. In addition, Al wires leading from the nanowire to qubit islands and nanowire to the ground plane were deposited. In the final step, the nanowire, qubit island and ground plane were electrically connected by creating ohmic contacts. To ensure a good contact $\mathrm{AlO}_{\mathrm{x}}$ on the Al wires and MBE Al was removed by Ar-milling prior to the deposition of a $\sim 250\,\mathrm{nm}$ thick Al layer.
		
		\section{Resonator fabrication and measurements}\label{app:Fab_res}
		
		Al test resonators were fabricated on the Si substrate after removing the entire GaAs and GaP buffer and deep etching into the Si substrate using the RIE steps described above. We also added patterned structures on the chip to measure the etch depth with a mechanical profilometer. The resonators were then defined by a global electron beam evaporation of $100\,\mathrm{nm}$ thick Al under high vacuum and subsequent selective etching of the Al using standard electron e-beam lithography and wet-etch techniques. Preceding the Al deposition, the device chip was dipped into buffered HF solution to remove surface oxides from the Si substrate.
				
\begin{figure}
			\includegraphics[width=1\columnwidth]{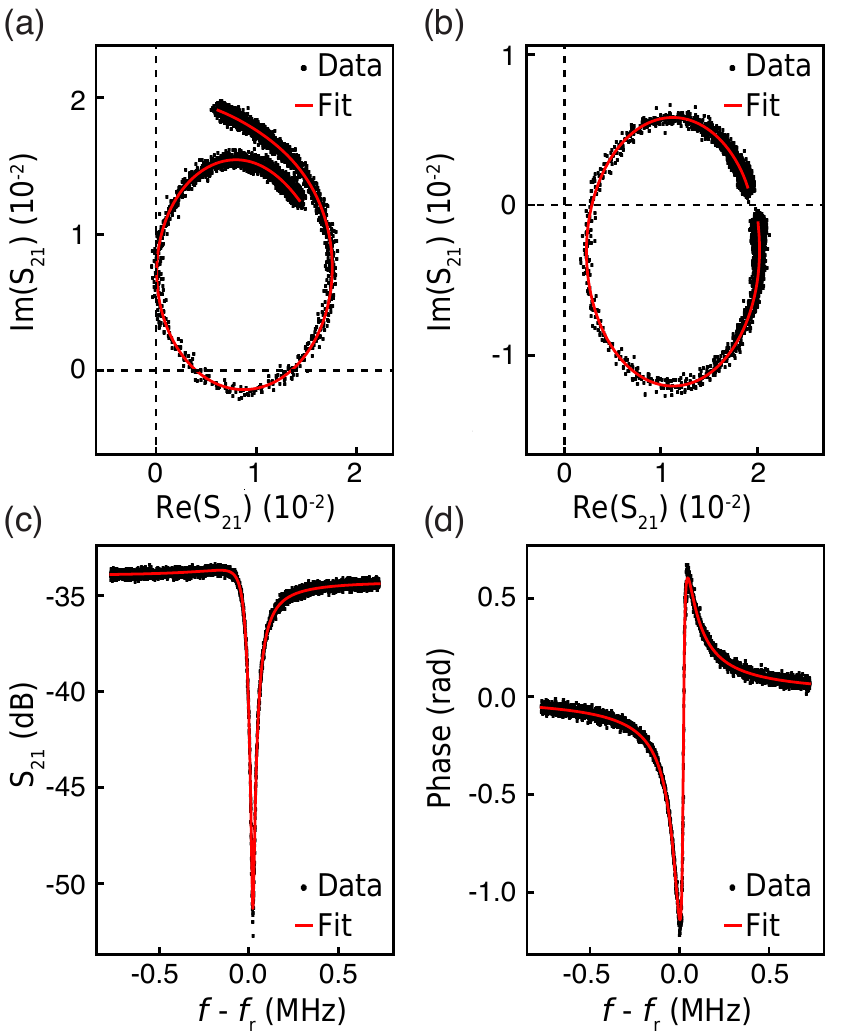}
			\setlength\belowcaptionskip{-16pt}
			\caption{\label{fig:Fig5}\textbf{Example data and fit of the resonator with an etch depth of $\mathbf{720}\,\mathrm{\mathbf{nm}}$ (see Fig.~\ref{fig:Fig1})} at $P_{\mathrm{on \text{-} chip}}=-149\,\mathrm{dBm}$. \textbf{(a)} Raw data (black) and fit (red) of the trajectory of $S_{21}$ in the complex plane. \textbf{(b)} $S_{21}$ data and fit after correcting for the cable delay \cite{Bruno2015, Probst2015}. \textbf{(c)} and \textbf{(d)} show the magnitude and phase of $S_{21}$ of the data in (a) with the corresponding fits.}
		\end{figure}

		For the extraction of the internal and external quality factors, $Q_{\mathrm{i}}$ and $Q_{\mathrm{ext}}$, we used the fit procedure described in Ref.~\cite{Bruno2015}. Fig.~\ref{fig:Fig5} shows an example fit in the few photon regime for a resonator made after etching the III-V layers and $720\,\mathrm{nm}$ of Si. Here, we used resonators for which $Q_\mathrm{ext} \approx Q_\mathrm{i}$ in the few photon regime. The data shown in Fig.~\ref{fig:Fig5} was taken at  $P_{\mathrm{on\text{-}chip}}=-149\,\mathrm{dBm}$, which corresponds to a mean photon number on chip of $\braket{n_{\mathrm{ph}}}\approx 1 $. The resonance frequency $f_{\mathrm{r}}$ of the resonator is at $f_{\mathrm{r}}=5.51\,\mathrm{GHz}$. From the fit we extracted $Q_\mathrm{i}\approx 3.6\times10^5$ and $Q_\mathrm{ext}\approx 57\times 10^3$. We note, that for extracting $Q_{\mathrm{i}}$, we use resonators for which the external quality factor $Q_{\mathrm{ext}}$ matches $Q_{\mathrm{i}}$ ($Q_{\mathrm{ext}} \approx Q_{\mathrm{i}}$) in the few photon regime.

		We measured three resonator devices with different etch times. The results are summarized in Tab.~\ref{Table_resonatorQs}. The quality factor for device 1 was extracted from a fit at $P_{\mathrm{on \text{-}chip}}=-125\,\mathrm{dBm}$, for device 2 at $P_{\mathrm{on \text{-} chip}}=-146\,\mathrm{dBm}$ and for device 3 at $P_{\mathrm{on \text{-} chip}}=-148\,\mathrm{dBm}$. 
		
			\begin{table}
			\begin{center}
				\setlength{\tabcolsep}{0.5em}
				\def\arraystretch{1.5}
				\begin{tabular}{|c| c | c | c | c | c |}
					\hline
					Sample  & Etch depth$\,$(nm) & $f_{\mathrm{r}}\,\mathrm{(GHZ)}$ & $Q_\mathrm{ext}\,(10^3)$ & $Q_\mathrm{i}\,(10^3)$ \\ \hline 
					1& 120 & 6.35 & 12  & 27 \\
					2& 390 & 5.46 & 20 & 102 \\
					3&  720 & 5.51 & 57 & 371 \\
					\hline
				\end{tabular}
			\end{center}
			\caption{\label{Table_resonatorQs} Three resonator devices fabricated on Si with different etching parameters and their extracted resonance frequency $f_{\mathrm{r}}$, external quality factor $Q_\mathrm{ext}$ and internal quality factor $Q_{\mathrm{i}}$.}
		\end{table}

		\section{Pulsed qubit measurements}\label{app:Measurement_qubit}
		
		All measurements in this work were performed with the qubit in the transmon limit ($E_{\mathrm{J}}\gg E_{\mathrm{C}}$) and in the dispersive regime $|f_{\mathrm{q}}-f_{\mathrm{r}}|\gg g/2\pi$, where $g/2\pi\approx80\,\mathrm{MHz}$ is the coupling between cavity and qubit island. In all experiments, the DC gate voltage $V_{\mathrm{G}}$ and the RF drive tone for the qubit are combined using a bias tee at the millikelvin stage of the dilution refrigerator and applied through the gate line. 
		
		We extracted the frequencies $f_{02}/2$ and $f_{01}$ (Fig.~ \ref{fig:Fig2}a) by numerically fitting two independent Lorentzian line shapes to the signal and using the position of peak maxima as transition frequencies. For all qubit measurements in this work, the pulse length $\tau$ corresponds to the length of the plateau of a Gaussian flat top pulse with a rise time of $2\sigma = 4\,\mathrm{ns}$. All measurements in this work were limited to pulse lengths of $\sim 30\,\mathrm{ns}$ due to the short dephasing time (see Section \ref{sec:Coherent_oscillations}). To increase signal-to-noise ratio and adjust for short qubit dephasing times as well as a low coupling between resonator and transmission line ($Q_{\mathrm{c}}\approx2~\text{x}~10^4$), the resonator was driven during the qubit manipulation and resonator readout (pulse sequence shown in Figs.~\ref{fig:Fig3}a) unless stated otherwise. Here, we estimate that the signal rise time of the resonator with $f_{\mathrm{res}} \approx 6\,\mathrm{GHz}$ is $\tau_{\mathrm{rise}} \approx Q_{\mathrm{c}}/2\pi f_{\mathrm{res}} \sim 480\,\mathrm{ns}$. A non-overlapping pulse sequence in which the resonator drive was applied when the signal integration started was used to measure the data in Fig.~\ref{fig:Fig3}b (Ramsey fringes). Measurements as shown in Fig.~\ref{fig:Fig3}a were used to calibrate $R^{\pi}_{\mathrm{X}}$ and $R^{\pi/2}_{\mathrm{X}}$ pulses that rotate the qubit state around the $x$-axis by $\pi$ and $\pi/2$. 
		
\begin{figure*}
			\includegraphics[width=2\columnwidth]{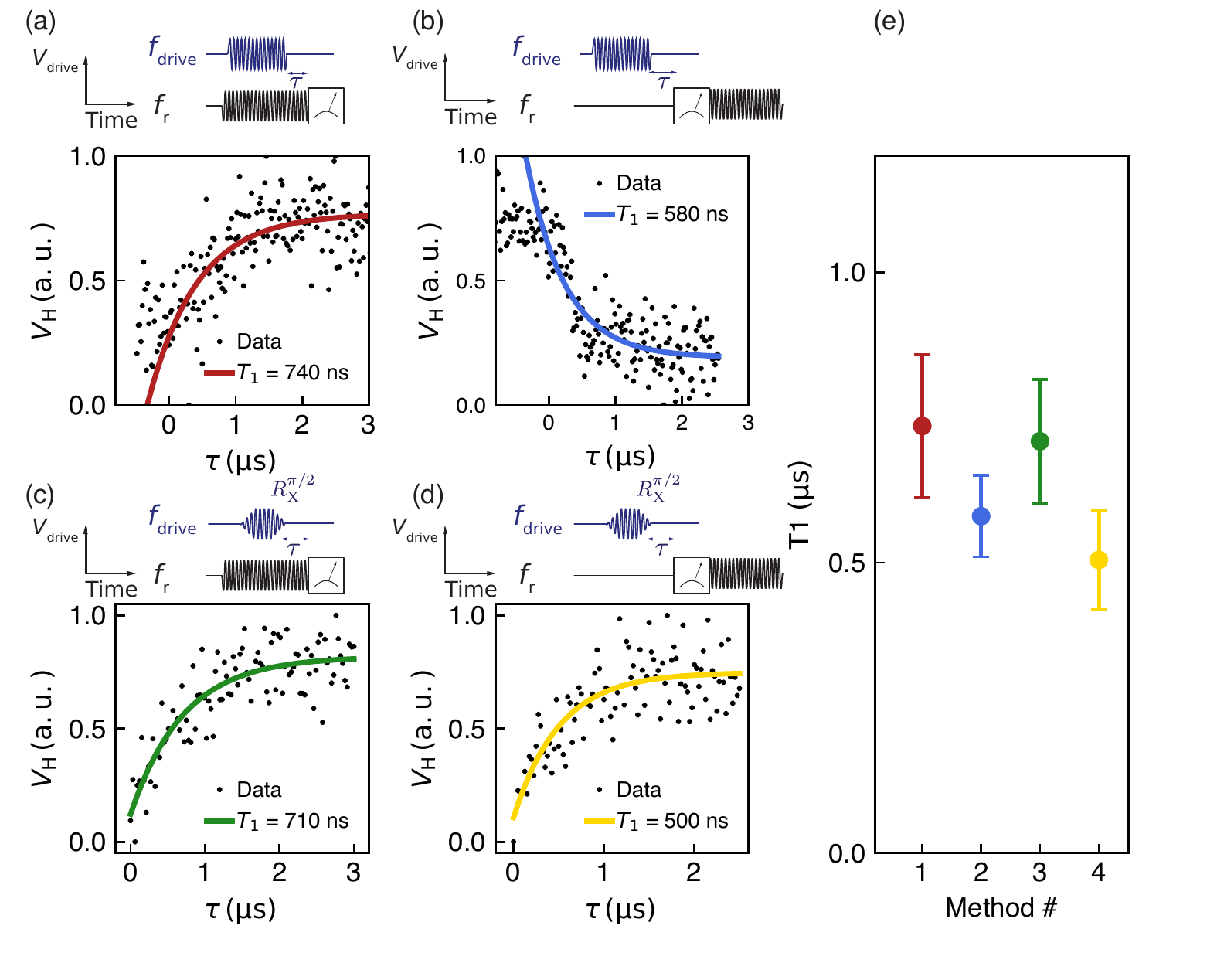}
			\caption{\label{fig:T1_methods} \textbf{Different measurement methods for extraction of relaxation time $T_1$}. \textbf{(a)} Readout and weak spectroscopy drive tone overlap. \textbf{(b)} The strong readout tone is applied after the spectroscopy drive tone. \textbf{(c)} A $\pi$-pulse is used to excite the qubit. This drive tone overlaps with the readout tone. \textbf{(d)} The readout tone is applied after $\pi$-pulse. \textbf{(e)}  Comparison of extracted $T_1$ values from datasets \textbf{(a)-(d)}.}
		\end{figure*}

	    We note that for our qubit the $T_2^{\ast}$ time ($\approx 15\,\mathrm{ns}$) is shorter than the ring-up time of the readout resonator ($\approx 480\,\mathrm{ns}$) and comparable to the minimum gate time to avoid state leakage, $\sim 1/(|\alpha|/h)$. In a standard measurement sequence including a $\pi$-pulse (see Figs.~\ref{fig:T1_methods}c and d) this could lead to leakage to higher order states. When driving we observe an additional slow oscillation in $V_{\mathrm{H}}$ (not shown) that we attribute to driving chip modes that are coupled to the qubit. The observed linear increase could be the onset of this slow oscillation. To test if the relaxation times are limited by leakage to higher order state we performed additional weak spectroscopy measurements. For these we used a weak spectroscopy pulse as drive tone as shown in Figs.~\ref{fig:T1_methods}a and b. Here, a low readout power was used to avoid an ac-Stark shift of the qubit caused by a high photon population in the resonator.  Fig.~\ref{fig:T1_methods}e compares the relaxation times extracted from the different datasets and their respective exponential fit as illustrated in Fig.~\ref{fig:T1_methods}a-d. The pulsed measurement results (Figs.~\ref{fig:T1_methods}c-d,) are in good agreement with the weak drive measurements.

		\section{Continuous drive $T_2^{\ast}$ measurement}\label{app:T2star}
		
\begin{figure*}
			\includegraphics[width=2\columnwidth]{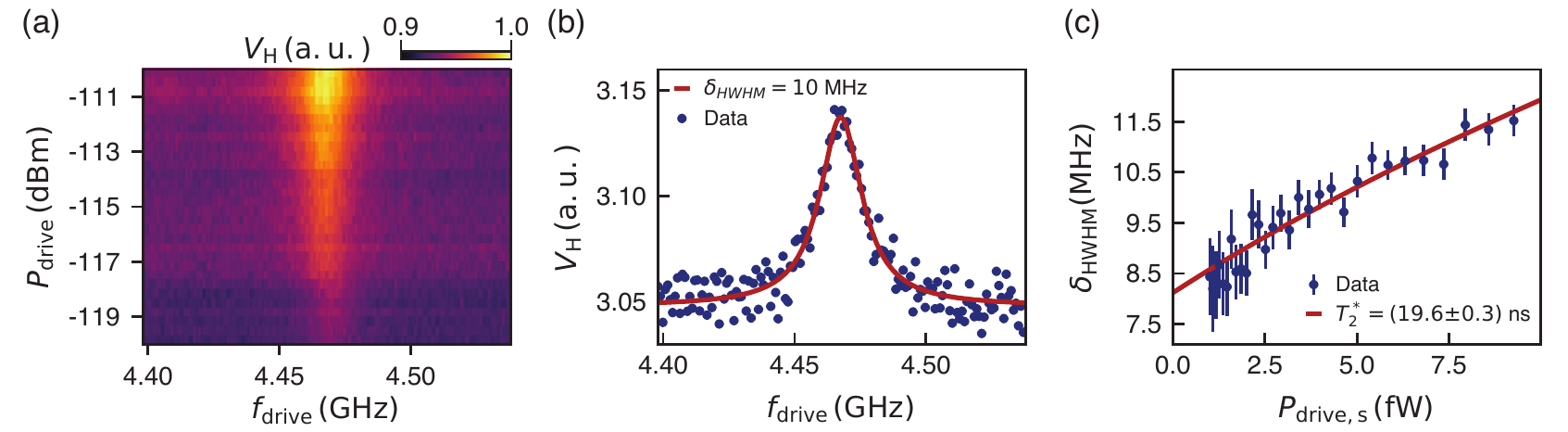}
			\caption{\label{fig:Fig_spec} \textbf{Spectroscopy data for estimation of dephasing time $T_2^{\ast}$}. \textbf{(a)} Spectroscopy signal as a function of qubit drive $P_{\mathrm{drive,s}}$ and drive frequency $f_{\mathrm{drive}}$. \textbf{(b)} Linecut from \textbf{(a)} at $P_{\mathrm{drive,s}}=-116.7\,\mathrm{dBm}$ with fit to Lorentzian lineshape (purple line). \textbf{(c)} Qubit linewidth $\delta_{\mathrm{HWHM}}$ vs. power $P_{\mathrm{drive,s}}$ with fit using $2\pi\delta_{\mathrm{HWHM}} = 1/T^{\prime}_{2}=(1/T_2^{\ast 2} + n_{\mathrm{s}}w_{\mathrm{vac}}^2T_{1}/T_2^{\ast})^{1/2}$. The decoherence time is estimated from the intersection of the fit with the y-axis.}
		\end{figure*}

		To estimate the dephasing time of the qubit, we follow the method described in Ref. \cite{Schuster2005}. Here, the linewidth of the qubit in spectroscopy can be described by the equation $2\pi\delta_{\mathrm{HWHM}} = 1/T^\prime_{2}=(1/T_2^{\ast 2} + n_{\mathrm{s}}w_{\mathrm{vac}}^2T_{1}/T_2^{\ast})^{1/2}$, where  $\delta_{\mathrm{HWHM}}$ is the half width at half maximum of the spectroscopy feature, $T^\prime_{2}$ is the dephasing time,  and $n_{\mathrm{s}}w_{\mathrm{vac}}^2$ is proportional to the drive power on-chip $P_{\mathrm{drive,s}}$ with $P_{\mathrm{drive,s}}=P_{\mathrm{drive}}-60\mathrm{dB}$ (60dB accounts for cryogenic attenuators, line attenuation and filtering as depicted in Fig.~\ref{fig:setup}) and $T_2^{\ast}$ is the dephasing time at zero drive power. Figure~\ref{fig:Fig_spec}a shows the spectroscopy signal as a function of $P_{\mathrm{drive,s}}$. At each value of $P_{\mathrm{drive,s}}$, we extract the linewidth by fitting to a Lorentzian line shape (Fig.~\ref{fig:Fig_spec}b). As shown in Fig.~\ref{fig:Fig_spec}c, $T_2^{\ast}$ is extracted from a fit to the power dependence of the linewidth. 
		
		\section{Gate voltage dependence}\label{app:LifetimeVsGate}
\begin{figure*}
			\includegraphics[width=2\columnwidth]{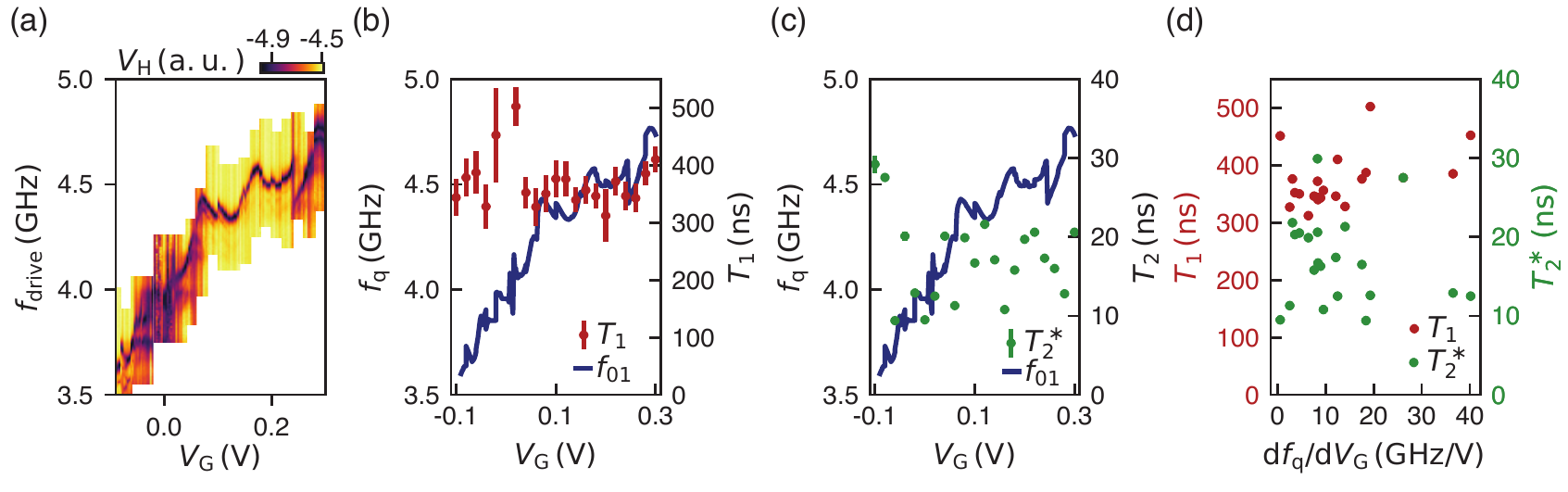}
			\caption{\label{fig:gate_dispersion} \textbf{Coherence times and $f_{\mathrm{q}}$ as a function of $V_{\mathrm{G}}$}. \textbf{(a)} Spectroscopy signal. \textbf{(b)} $f_{\mathrm{q}}$ as a function of $V_{\mathrm{G}}$ and $T_1$ for 20 different gate voltages. \textbf{(c)} Same as (b) but for $T_2^{\ast}$. \textbf{(d)} Coherence times $T_1$ and $T_2^{\ast}$ as a function of gate dispersion $\mathrm{d}f_{\mathrm{q}}/\mathrm{d}V_{\mathrm{G}}$.}
		\end{figure*}

\begin{figure*}
		\includegraphics[width=2\columnwidth]{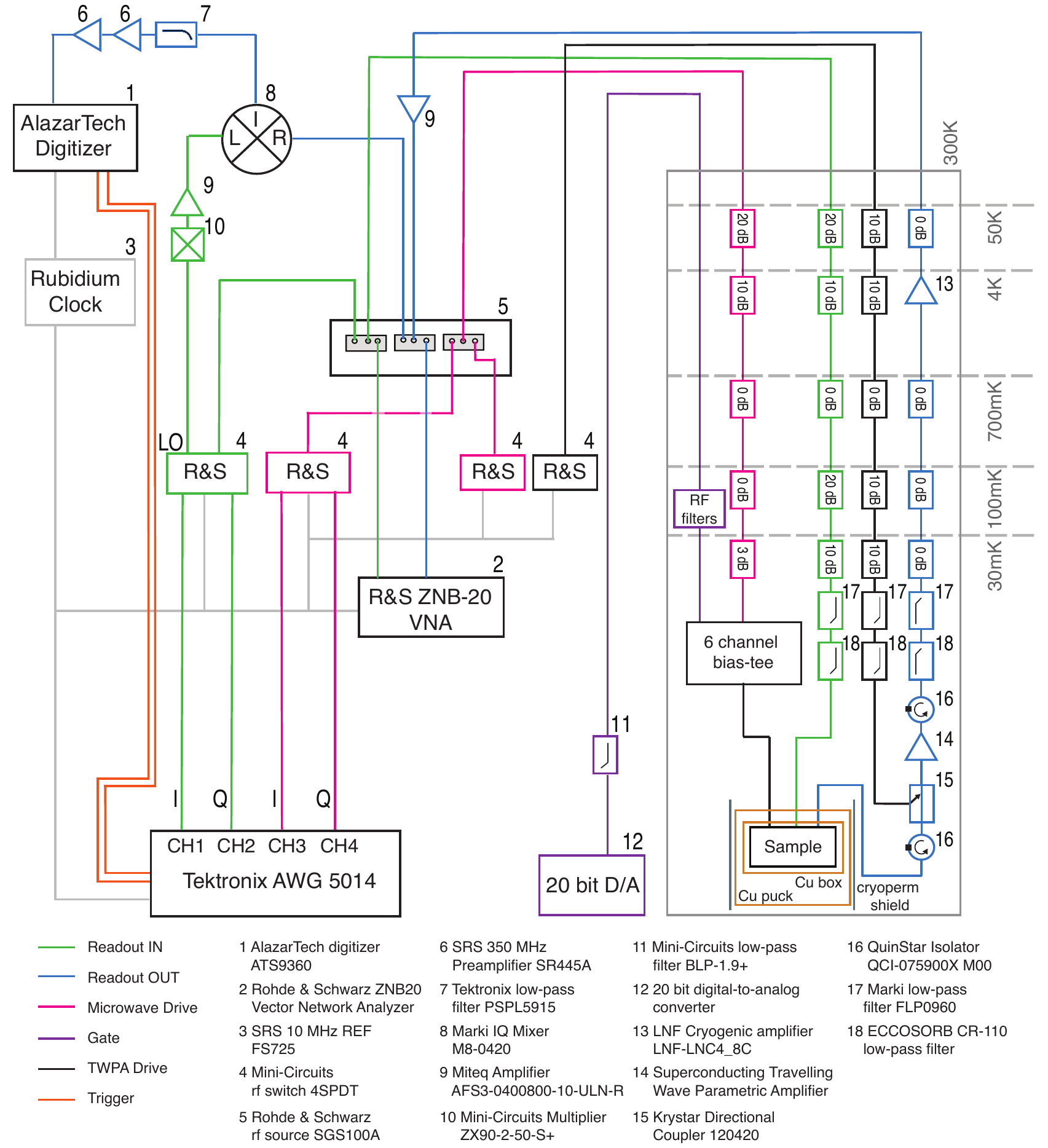}
		\setlength\belowcaptionskip{-8pt}
		\caption{\label{fig:setup} \textbf{Schematic of the experimental setup used for the experiments.} The readout resonator can be driven by a vector network analyzer (VNA) or by a signal from a rf source that is modulated by an arbitrary waveform generator (AWG) (blue line). The amplified output signal can be read out either by the VNA or undergo down conversion by mixing with a reference signal. Microwave equipment is synchronized using a 10MHz clock reference. To tune the chemical potential of the SAG nanowire junction, a DC line (purple) is connected to the sample. The DC signal merges with the (pink) RF signal used for driving the qubit in a bias tee channel.}
		\end{figure*}

		We performed $T_1$ measurements and spectroscopy sweeps to extract $T_2^{\ast}$ from the line width at different gate voltages. In Fig.~\ref{fig:gate_dispersion}a we show the spectroscopy signal for gate voltages $-0.09\,\mathrm{V} < V_{\mathrm{G}}< 0.3\,\mathrm{V}$. In this range $3.6\,\mathrm{GHz} <f_{\mathrm{q}}(V_{\mathrm{G}})<4.8\,\mathrm{GHz}$. In Figs.~\ref{fig:gate_dispersion}b, c the relaxation time $T_1$ and dephasing time $T_2^{\ast}$ are shown for 20 different values of $V_{\mathrm{G}}$ and the qubit frequency $f_{\mathrm{q}}(V_{\mathrm{G}})$ which we extracted from the spectroscopy data. Figure~\ref{fig:gate_dispersion}d shows $T_1$ and $T_2^{\ast}$ as a function of gate dispersion $\mathrm{d}f_{\mathrm{q}}/\mathrm{d}V_{\mathrm{G}}$ calculated numerically from $f_{\mathrm{q}}(V_{\mathrm{G}})$. As $T_1$ and $T_2^{\ast}$ show no correlation with qubit frequency $f_{\mathrm{q}}(V_{\mathrm{G}})$ we conclude that the qubit coherence times are not limited by gate noise.
		
		\section{Potential fabrication improvements}\label{app:roadmap}
		The mesa could be placed further away from strong electric fields generated by the qubit island and the mesa size could be reduced. This could be achieved by using a different dry etch chemistry. An isotropic wet etch could be used to create a concave mesa shape, leading to an increased distance and thus smaller coupling between the mesa and the residual circuit. A possible improvement for future devices is the definition of the mesa before the selective area growth, leading to a reduced mesa size, set by the diffusion length of particles during the MBE growth, and enabling the fabrication of the readout circuit and qubit island using MBE Al. Side gates could be used instead of top gates to remove gate dielectric from the process flow. The growth dielectric around nanowires could be removed by means of HF vapor etching or design changes. In order to decouple qubits from the lossy mesa stack, thicker PMMA bridges could be used. 
		
		\section{Additional sources for decoherence}\label{app:decoherence}
		We can exclude setup related and measurement related dephasing mechanisms as limiting factors for the short dephasing times. Our system was calibrated with metallic fixed-frequency transmon devices, where $T_2^{\ast}\approx 10\,\mathrm{\micro s}$ was measured, and we used the same sample packaging as in Refs. \cite{Larsen2015, Casparis2018}, where $T_2^{\ast}\approx900\,\mathrm{ns}$ and $T_2^{\ast}\approx400\,\mathrm{ns}$ were reported. Further, no correlation between the $T_2^{\ast}$ and the gate dispersion $\mathrm{d}f_{\mathrm{q}}/\mathrm{d}V_{\mathrm{G}}$ is observed, indicating lifetimes are not limited by decay through the gate line or gate noise (Appendix \ref{app:LifetimeVsGate}). Here we note that the Rabi frequencies in these measurements are comparable to the qubit anharmonicity $|\alpha|/h \sim 120\,\mathrm{MHz}$, potentially leading to leakage to higher order states and inducing dephasing \cite{Peterer2015}. To reduce the potential impact of this effect, Gaussian flat-top pulses were used to avoid driving higher order transition of the qubit system. To avoid additional dephasing due to photon number fluctuations we chose a low cavity readout power. Further, we confirmed that this overlapping pulse sequence yields the same lifetimes as alternative measurement techniques where we used non-overlapping pulses or prepared the qubit in a well-defined mixed state using long and weak spectroscopy pulses \cite{Mergenthaler2021} (see Appendix \ref{app:Measurement_qubit}).

	\section{Setup}

	The presented measurements were acquired in a cryofree dilution refrigerator with a base temperature of $\sim$ 30mK. Fig.~\ref{fig:setup} shows a detailed schematic of the experimental setup. An RF switch is used to route signals to the sample that are either coming from the vector network analyzer (VNA) or from AWG modulated rf sources. The input signals are attenuated and filtered before reaching the sample. For measurements the RF switch can direct the readout signal back to the VNA or to the demodulation circuit. The latter consists of a mixer for down conversion of the output signal to an intermediate frequency by mixing it with a reference tone. The signal is then filtered and amplified before it is finally digitized and digitally downconverted with the AlazarTech Digitizer. 
	The microwave drive tone is generated by an additional R$\&$S microwave source and is applied via its own  drive line that merges with the gate line in a bias tee before reaching the sample.
	For synchronisation of all instruments an SR FS725 10 MHz clock reference is used.
	
	\end{appendix}


%

\end{document}